\documentclass[floatfix,prb,aps]{revtex4}
\usepackage{graphicx}
\usepackage{psfrag}
\usepackage{epstopdf}

\newcommand{\ket}[1]{\left|#1\right\rangle}

\begin{document}

\title{\bf Readout methods and devices for Josephson-junction-based solid-state qubits.}
\author{G. Johansson, L. Tornberg, V.S. Shumeiko and G. Wendin}
\affiliation{Department of Microtechnology and Nanoscience - MC2,
Chalmers University of Technology, \\SE-41296 Gothenburg, Sweden}

\date{\today}

\begin{abstract}
We discuss the current situation concerning measurement and
readout of Josephson-junction based qubits. In particular we focus
attention of dispersive low-dissipation techniques involving
reflection of radiation from an oscillator circuit coupled to a
qubit, allowing single-shot determination of the state of the
qubit. In particular we develop a formalism describing a charge
qubit read out  by measuring its effective (quantum) capacitance.
To exemplify, we also give explicit formulas for the readout time.
\end{abstract}

\maketitle

\section{Introduction}

Nanotechnology is considered promising for fabrication of scalable
solid-state electronics for quantum computers
\cite{EsteveVion2004,Elzerman2005,WendinShumeiko2005}. However,
progress towards solid-state quantum computing will critically
depend on the development of measurement schemes and readout
devices that, on demand, can determine the state of individual
qubits in a fraction of the coherence time, but which otherwise do
not disturb the qubit system. In quantum optics, efficient
measurement techniques have been developed during the last thirty
years based on laser-atom interactions and recently implemented in
e.g. ion traps
\cite{Roos2004,Riebe2004,Barrett2004,Chiaverini2004,Leibfried2005,Haffner2005}.

The corresponding work for solid state systems effectively started only about
ten years ago, and is currently exploring various paths. A problem is that
there is no general device for operation and readout, like a laser, but
rather a multitude of implementations of measurements of charge, spin,
magnetic flux and charge current that must be adapted to the specific qubits
to be studied. Therefore, the qubit readout technology must be developed in
intimate connection with the qubits for characterization and control of
coherence properties. This is a painstakingly slow process which, however,
cannot be circumvented, because it is essential in many respects. In
particular, it is an important tool for determining the coherence properties
of the qubits. Moreover, the technology not only concerns qubit readout
devices, but also involves quantum oscillators for storing and transmitting
information and for coupling qubits.

Interestingly enough, quantum-optical methods are now being
applied to solid-state qubit systems, using microwaves for
operating and reading out qubits, and oscillator circuits and
transmission lines for coupling qubits, introducing cavity-QED in
solid-state systems \cite{Blais2004,Rau2004,Wallraff2004}. This
may turn out to be a major road on the Road Map for quantum
coherent systems ("quantum computers")
\cite{EU-QIPC2005,US-QC2005}, and will be at the focus of the
present paper. In particular we will describe some practical schemes
for reflecting microwaves from an oscillator circuit, the phase shift
measuring the changes in charge \cite{Sillanpaa2005,Duty2005}
or magnetic flux \cite{Zorin2002,Ilichev2003,Sillanpaa2004,Lupascu2005,Bertet2005}
induced by a qubit, allowing to distinguish between the different
states $\ket{0}$ and $\ket{1}$ of the qubit.

\section{Measurement of quantum information and qubit readout}
\label{SectII}

\subsection{Introduction}

The ultimate objective of a qubit readout device is to
distinguish the {\em eigenstates} of a qubit in a single
measurement "without destroying the qubit", a so called
"single-shot" quantum non-demolition (QND) projective measurement.
This objective is essential for several reasons: state preparation
for computation, readout for error correction during the
calculation, and readout of results at the end of the calculation.
Strictly speaking, the QND property is only needed if the qubit
must be left in an eigenstate after the readout. In a broader
sense, readout of a specific qubit must of course not demolish
other qubits in the system.

Note that one cannot "read out the {\em state} of a qubit" in a
single measurement - this is prohibited by quantum mechanics. The
effect of a single ideal measurement on a qubit in a general
superposition $a \ket{0}+b\ket{1}$ is to leave the qubit in one of
the states $\ket{0}$ or $\ket{1}$, which carries no information
about the amplitudes $a$ or $b$. It takes repeated measurements on
a large number of replicas of the quantum state to characterize
the state of the qubit  - "quantum tomography". This is the
procedure to collect the statistics for expectation values.

The measurement connects the qubit with the open system of the
detector, which collapses the combined system of qubit and
measurement device to one of its common eigenstates. If the
coupling between the qubit and the detector is weak, the
eigenstates are approximately those of the qubit. In general
however, one must consider the eigenstates of the total
qubit-detector system and manipulate gate voltages and fluxes such
that the readout measurement is performed in a convenient energy
eigenbasis (see e.g. Refs. \onlinecite{MakhlinRMP2001} and
\onlinecite{Wilhelm2003}).

\subsection{Survey of readout methods for JJ-based qubits}
Here we will provide a brief recapitulation of the "history" of
readout of Josephson-junction (JJ) based quantum circuits and qubits. For an extensive
discussion of JJ-based qubit circuits, see the recent review by Wendin and Shumeiko\cite{WendinShumeiko2005}.

Figure 1 shows general designs for the charge and flux qubits and
with oscillator-type readout circuits. The Single Cooper-pair Box (SCB)
(similar to Fig. 1, left) is described by the Hamiltonian
\begin{equation}\label{HQcpb}
\hat H_{} = E_C ( n - {n_g})^2 - E_J\cos\phi \;
\end{equation}
and the rf-SQUID (similar to the flux qubit  in Fig. 1, right; see Section VI) by the Hamiltonian
\begin{equation}\label{HQsquid}
\hat H =  E_C\, n^2 - E_J\,\cos\phi + E_L {(\phi-\phi_e)^2\over 2};
\end{equation}
where $E_C$ is the charging energy of the SCB island,  $E_J$ is the Josephson energy due to Cooper-pair tunneling between the superconducting electrode and island across the JJ,
and  $E_L$ is the inductive energy of the superconducting loop.
$n$ and $\phi$ are operators for the induced charge and the phase of the (effective) Josephson junction, and $n_g$ is the induced charge controlled by the external bias voltage $V_g$.
Both qubits are represented by the generic  2-level  Hamiltonian
\begin{eqnarray}
\hat{H} = -{1\over2}({\epsilon}\;\sigma_z + \Delta \;\sigma_x)
\end{eqnarray}
where $\sigma_z$ and $ \sigma_x$ are the usual Pauli matrices.

\begin{figure}[!ht]
\label{qubits}
\includegraphics[width=5cm]{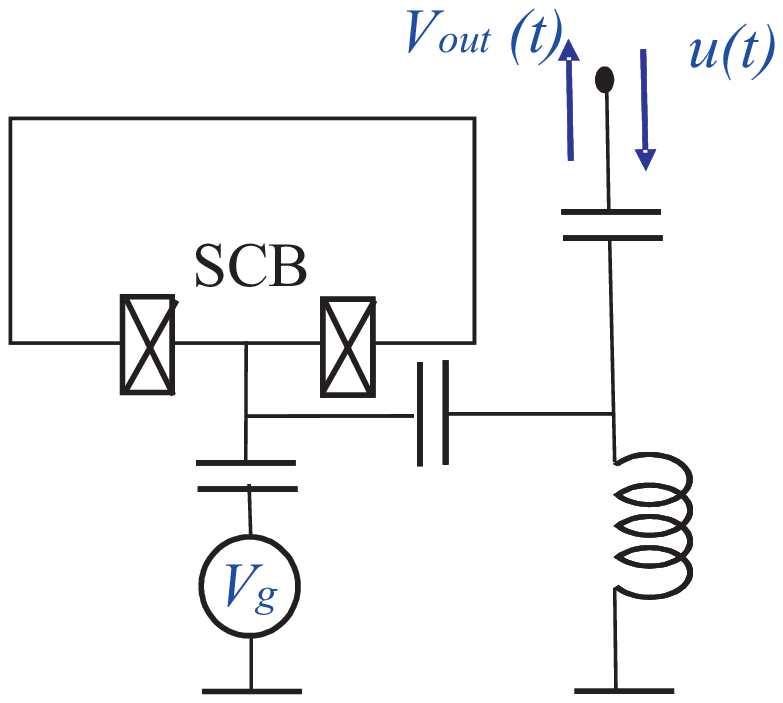} $\; \; \;$
\includegraphics[width=4.5cm]{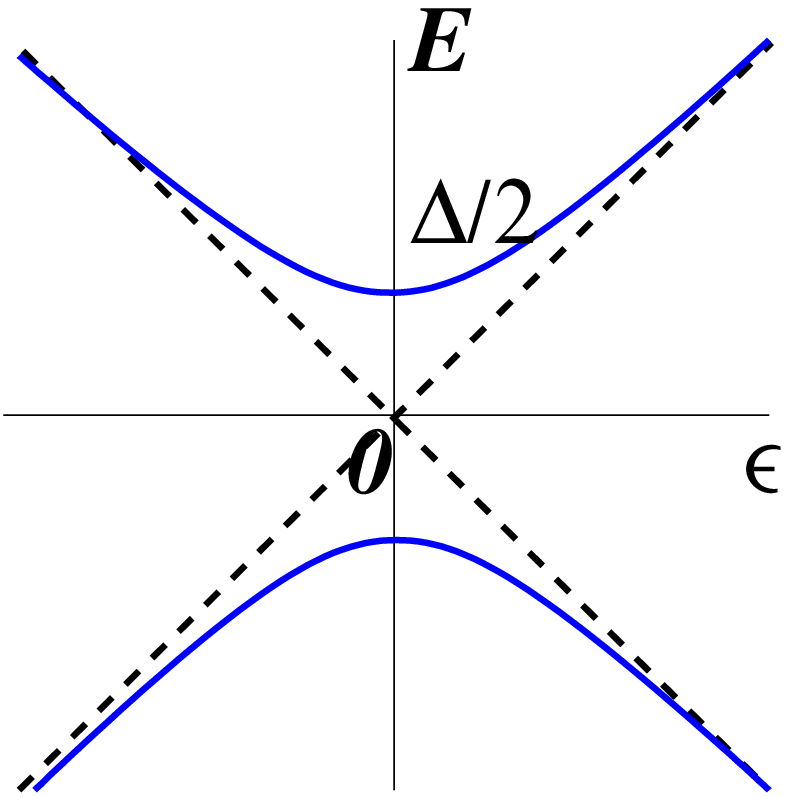} $\; \; \; \; \; \; $
\includegraphics[width=5cm]{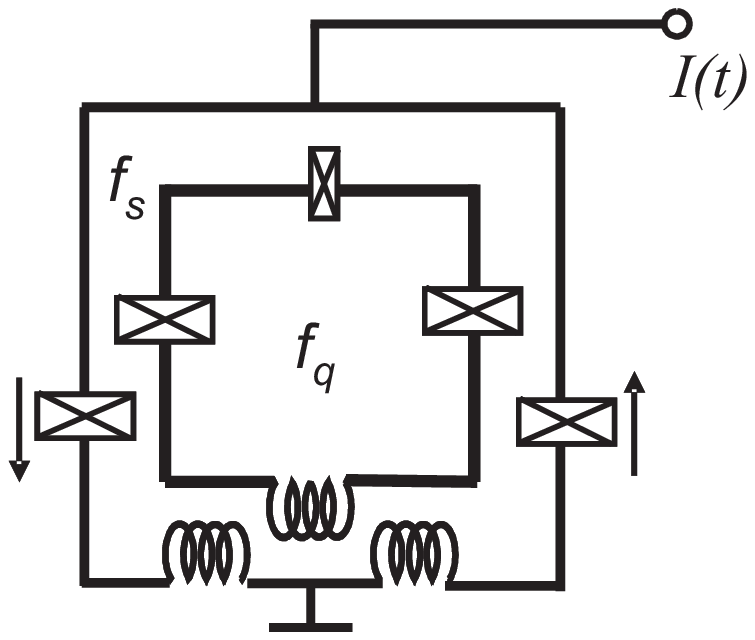}
\caption{Circuit diagrams and 2-level energy spectrum of two basic
JJ-qubit designs: the SCB charge qubit with LC-oscillator readout
(left), and persistent-current flux qubit with SQUID oscillator
readout (right). For the charge qubit, the control variable
$\epsilon$ on the horizontal axis of the energy spectrum (middle)
represents the external gate voltage (induced charge), and the
splitting is given by the Josephson tunneling energy mixing the
charge states. For the flux qubit, the variable $\epsilon$
represents the external magnetic flux. In both cases, the energy
of the qubit can be "tuned" and the working point controlled.
Away from the origin (asymptotically) the levels represent pure
charge states (zero $\ket{0}$ or one $\ket{1}$ Cooper pair on the
SCB island) or pure flux states (left $\ket{0}$ or right $\ket{1}$
rotating currents in the SQUID ring). }
\end{figure}

In the original experiment of Nakamura et al. \cite{Nakamura1999},
demonstrating coherent oscillation of the charge qubit  2-level system (Fig.
1, left), the readout was implemented simply by a control dc-pulse on the
charge gate, moving the working point far away from the origin so that the
upper $\ket{1}$ level ended up above the gap edge of a superconducting lead
connected to the SCB island via a tunnel junction.
As a result, a Cooper pair on the upper level $\ket{1}$ would immediately
decay into the external lead as two quasiparticles, creating a normal
electron current. Repeating the measurement at a high rate created a
detectable current proportional to the occupation of the upper state
$\ket{1}$, revealing the oscillations. Since the SCB is permanently connected
to the environment via a tunnel junction, it seemed plausible at the time
that this might be the reason for the short coherence time, $\sim$ 2-3 $ns$.

This focussed the interest on more advanced readout devices. A remedy could
be to use a charge measuring device that was only capacitively coupled to the
SCB island and could be turned on and off by an external voltage pulse.
Delsing and coworkers \cite{Duty2004} therefore developed an rf-SET
(radio frequency single-electron transistor) readout
\cite{Schoelkopf1998} for the charge qubit, and successfully detected free
oscillations and studied the detailed behaviour of relaxation and dephasing
\cite{Duty2004}. The result showed that the coherence time was confined to
below 10 $ns$ and seemed limited by relaxation effects. Moreover, subsequent
experiments by the NEC group \cite{Astafiev2004}, implementing more advanced
readout concepts, storing the emitted pair of quasiparticles on a
superconducting island, and reading the charge with a superconducting SET,
made no big change. All in all, the status seems to be that the coherence
time of the circuit is severely limited by intrinsic charge fluctuation
processes (noise) in the substrate, or in the tunnel barriers, or by
transients due to the pulsed operation of the qubit.

Alternatively one could create a new type of charge qubit by connecting the
Cooper  Pair Box island to two JJ tunnel junctions, creating a Single Cooper
Pair Transistor (SCT). This could be probed via charge \cite{Duty2005} or
current \cite{Vion2002,Vion2003,Corlevi2005,Sjostrand2005} measurements.
These experiments can be designed either as threshold detection measurements
or as microwave-reflection measurements. The reflection measurements with
phase-shift detection will be the main theme of this paper.

Moreover, there is the persistent-current  flux qubit, based on  a quantum version
of the RF-SQUID
\cite{vdWal2000,Friedman2000,Chiorescu2003,Chiorescu2004} coupled
to a measurement dc-SQUID. This measurement SQUID be operated
either as a current threshold detector
\cite{Chiorescu2003,Chiorescu2004} or as resonance circuit
reflecting and phase-shifting microwave radiation
\cite{Lupascu2005,Bertet2005}. We will briefly describe the microwave
reflection measurement also in this case in sect. VI.

\section{Charge measurements}
The most straightforward way to read out a charge qubit is to
measure its charge. As discussed above, to obtain a high fidelity
read out one should perform a measurement in the qubit eigenbasis.
This removes the possibility for the qubit to switch its state
during the measurement. When the measurement basis is fixed, as
determined by a charge measurement, we need to bias the qubit
where the charge basis is the eigenbasis. For the superconducting
charge qubit this implies a complete quenching of the Josephson
energy, while for quantum dot charge qubits one needs raise the
tunnel barrier between the dots. Having quenched the transitions
between the charge states the fidelity will in theory be perfect,
irrespectively of the measurement speed. In reality there is
always some remaining transition/relaxation channel open which
implies the need of a fast read-out. Fast read-out is also
mandatory for implementing an error correcting algorithm, where
the read out and correction should be performed on a time-scale
set by decoherence of the other qubits.

\subsection{The radio-frequency single-electron transistor}
The state of the art charge measurement device is the
radio-frequency single-electron transistor (rf-SET)
\cite{Schoelkopf1998} with an experimentally measured sensitivity
of $\delta q=3.2 \cdot 10^{-6}$
e/$\sqrt{\rm{Hz}}$\cite{AassimeAPL}. The measurement time
$t_{ms}$ needed to separate two states with a charge difference
$\Delta Q$ is $t_{ms}=(2\delta q/\Delta Q)^2$, indicating the
possibility of detecting a charge difference of one percent of the
electron charge ($\Delta Q=0.01$ e) in half a microsecond.
\begin{figure}[!ht]
\includegraphics[width=5cm]{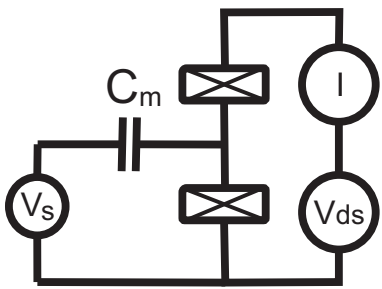}
\caption{A single-electron transistor: a small metallic island
connected to source and drain leads through tunnel junctions. The
signal voltage $V_s$ induces a charge $q_m$ on the measurement
capacitance $C_m$. When the system is at the limit of being
Coulomb blockaded, a small change of $q_m$ will have a large
effect on the current $I$ through the SET.}\label{SETFig}
\end{figure}
\begin{figure}[!ht]
\includegraphics[width=10cm]{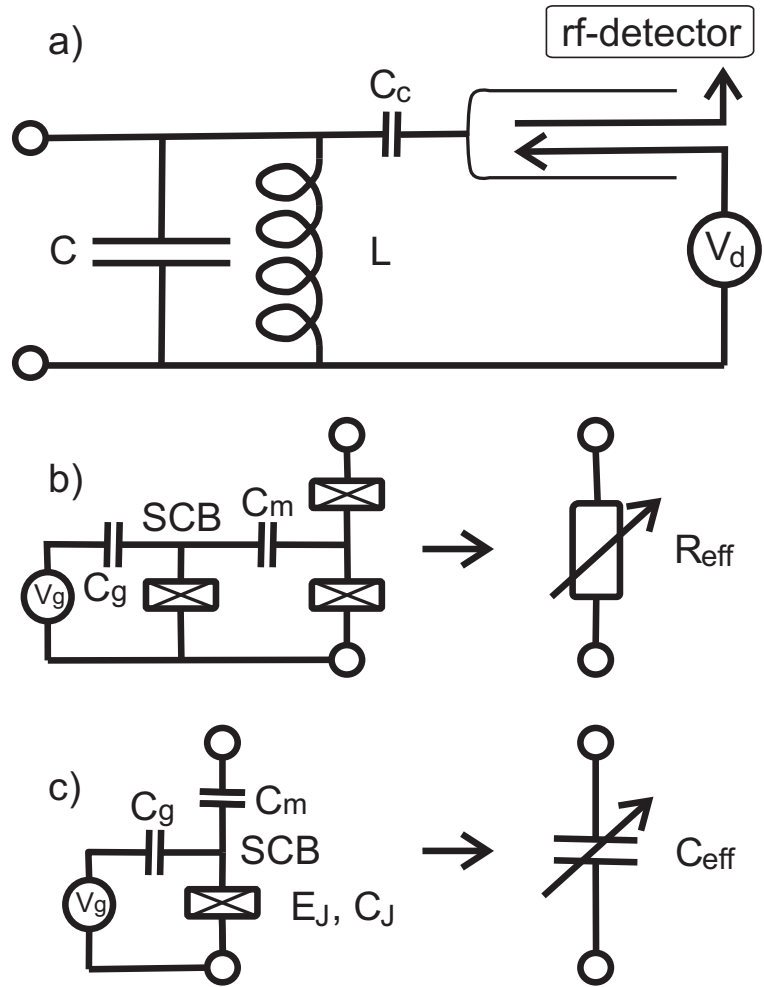}
\caption{Resonant circuits for read-out: a) A lumped element
LC-oscillator coupled to a driving source and a radio-frequency
detector through a transmission line. b) The radio-frequency
single-electron transistor measuring the charge of a charge qubit
(SCB). The current through the SET determines the dissipation in
the resonant circuit. The dissipation is determined by measuring
the amplitude of the reflected signal. c) Setup for measuring the
quantum capacitance of the charge qubit. The qubit capacitance
influences the resonance frequency of the oscillator. The
capacitance is measured by determining the phase-shift of the
reflected signal. } \label{ResonantCircuitsFig}
\end{figure}

The single-electron transistor (SET) consists of a small metallic
island connected to source and drain leads through tunnel
junctions (see Fig. 2). Applying a source-drain voltage
the current through the SET depends critically on the charge
induced on the gate capacitance. The charge on the gate
capacitance is determined by measuring the current. The SET is in
itself a sensitive electrometer but suffers from low operating
speed, which in addition to being a drawback on its own makes it
sensitive to the low frequency charge fluctuations. The rf-SET is
realized by embedding an SET in a resonant circuit (see
Fig. Fig. 3 a,b). The main source of
dissipation in the oscillator is current flowing through the SET,
and for small amplitude oscillations we may replace the SET by its
effective (differential) resistance. The oscillator is excited by
sending down a radio-frequency signal on resonance. The
dissipation is determined by measuring the amplitude of the
reflected signal. This way of operating the SET increases its
operating speed and sensitivity significantly.

\subsection{Single-shot read-out}
The high sensitivity can be used to couple the rf-SET weakly to
the charge qubit, reducing the back-action in the off-state. In
practice it is impossible to switch off the interaction completely,
i.e. the qubit eigenbasis is not exactly the charge basis,
and there is some unavoidable mixing of the charge states .
A careful investigation of the rf-SET coupled to
a superconducting charge qubit shows that single-shot read-out
with a very high fidelity is still possible in practice
\cite{AassimePRL,JohanssonPRL,KaeckPRB}. To optimize the
fidelity one should bias the charge qubit as far away from the
degeneracy point as possible, in order to minimize the effect of
any residual Josephson coupling.

\section{Capacitance measurements \-- a quasiclassical description}
The low-frequency charge fluctuations present in all realizations
of superconducting charge qubits so far
\cite{AstafievPRL2004,SaclayKarlsruhePRB} promotes the use of
schemes where the charge qubits are operated at the charge
degeneracy point around the origin in Fig. 1 (middle panel).
Here the qubit eigenstates have equal average
charge and thus they are shielded from charge fluctuations. To use
rf-SET read-out described in the previous section one needs to
quickly shift the qubit far away from degeneracy simultaneously with
switching-on the measurement. Although this timing is far from
impossible, the present trend is to use schemes where the qubit
remains at charge degeneracy also during read-out in order to minimize
decoherence. This can be
achieved by measuring the effective capacitance of the charge
qubit, as we describe below.

\subsection{Quantum capacitance of a single-Cooper-pair box}
The quantum capacitance of the Cooper-pair
box\cite{ButtikerQCPRB,BouchiatPhysicaScripta} is related to the
parametric capacitance of small Josephson
junctions\cite{WidomJLTP,AverinJETP,LikharevJLTP} which is a dual
to the Josephson inductance. The origin of the quantum capacitance
of a single-Cooper-pair box (SCB) can be understood as follows.
Assume that we put a constant voltage $V_m$ on the measurement
capacitance of the SCB, i.e. we put a voltage source between the
open circles in Fig.~\ref{ResonantCircuitsFig}c. The amount of
charge on the measurement capacitance $q^{g/e}_m(V_m,V_g)$ will be
a nonlinear function of the voltage $V_m$ as well as the gate
voltage $V_g$ and whether the qubit is in the ground or excited
state. We may define an effective (differential) capacitance
\begin{equation}
C_{{\rm eff}}^{g/e}(V_m,V_g)=\frac{\partial}{\partial V_m}
q^{g/e}_m(V_m,V_g) ,
\end{equation}
as seen from the measurement circuitry. Away from the charge
degeneracy points of the SCB no charge will float across the
Josephson junction and the effective capacitance is simply the
geometric capacitance $C_{{\rm geom}}=C_J C_m/(C_J+C_m)$ of the
Josephson junction capacitance $C_J$ and the measurement
capacitance $C_m$ in series. Around the charge degeneracy point a
change of voltage will induce a shift of a Cooper-pair across the
Josephson junction. For a voltage changing slowly on the timescale
of the inverse qubit gap $\hbar E_J^{-1}$ this charge
redistribution is dissipationless. This contribution to the
effective capacitance, which depends on the qubit state, we call
the quantum capacitance $C_Q^{g/e}$. From the SCB Hamiltonian (see
e.g. Eq.~(\ref{eq:Hq4})) it is straightforward to show
\begin{equation}
C_{{\rm eff}}^{g/e}(V_m,V_g)=\frac{C_J C_m}{C_J+C_m} \pm
\frac{C_m^2}{C_\Sigma} \frac{E_J^2 E_Q}{\left(E_Q^2[1-2(n_g +
n_m)]^2+E_J^2\right)^{3/2}} = C_{\rm geom}+C_Q(n_g+n_m),
\end{equation}
where $C_\Sigma=C_J+C_m+C_g$ is the total island capacitance,
$E_Q=2e^2/C_\Sigma$ is the Coulomb energy of a Cooper-pair, and
$n_{g/m}=C_{g/m}V_{g/m}/2e$ are the induced number of Cooper-pairs
on the gate and measurements capacitances respectively. We note
that the quantum capacitance is positive in the ground state and
negative in the excited state, as illustrated in Fig.  4. The absolute value is largest at
the charge degeneracy $n_g+n_m=0.5$
\begin{equation}
\label{eq:QuantCap} C_Q^{\rm
max}=\frac{C_m^2}{C_\Sigma}\frac{E_Q}{E_J}.
\end{equation}
The quantum capacitance can be an order of magnitude larger than
the geometric capacitance for realistic parameters.
\begin{figure}[!ht]
\label{CQPlotFig}
\includegraphics[width=10cm]{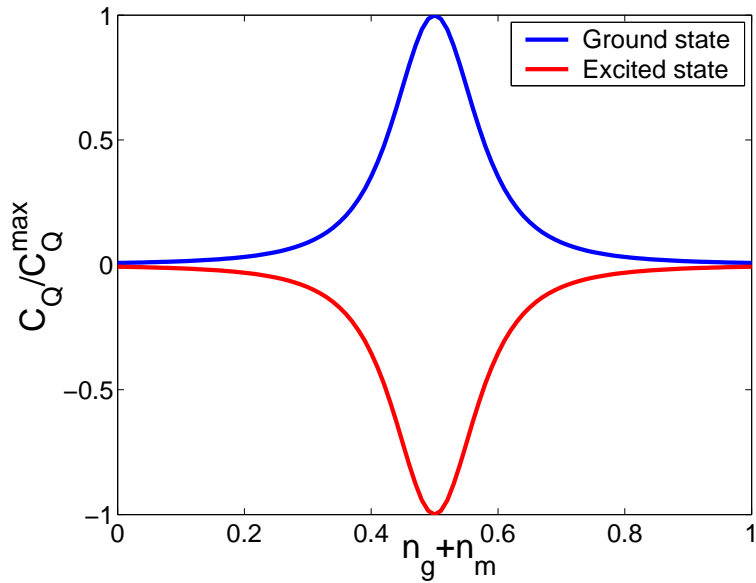}
\caption{The quantum capacitance of a single-Cooper-pair box with
$E_J/E_Q=0.2$. Here normalized by the maximum value $C_Q^{\rm
max}=\displaystyle\frac{C_m^2}{C_\Sigma}\frac{E_Q}{E_J}$. }
\end{figure}
By inserting the Cooper-pair box in a resonant LC-circuit and
detect its influence on the resonance frequency, the quantum
capacitance was recently measured by two different
groups\cite{Sillanpaa2005,Duty2005}. The measurement setup is
similar to the setup that would be used for qubit read-out and is
analyzed below.

\subsection{Read-out by measuring the quantum capacitance}
At the charge degeneracy point the effective capacitance of the
SCB in the ground and excited state differs by $2 C_Q^{\rm max}$.
Imbedding the SCB in a resonant circuit as shown in
Fig.~\ref{ResonantCircuitsFig} a) and c) we can detect the
corresponding change in the oscillators resonance frequency
$\omega^{g/e}_0=1/\sqrt{L (C\pm C_Q^{\rm max})}=\omega_0(1\mp
C_Q^{\rm max}/2C)$, where $\omega_0=1/\sqrt{LC}$ is the bare
resonance frequency. The voltage reflection amplitude
$\Gamma(\omega)=V_{out}(\omega)/V_d(\omega)$ seen from the driving
side of the transmission line can for a high quality oscillator be
written
\begin{equation}
\label{ClassicalVoltageReflectionAmplitude}
 \Gamma(\omega)=
\frac{1+2iQ\displaystyle\frac{(\omega-\omega_0)}{\omega_0}}
{1-2iQ\displaystyle\frac{(\omega-\omega_0)}{\omega_0}}=e^{i\varphi_r}, \ {\rm
where} \ \
\varphi_r=2\arctan{\left(\frac{2Q(\omega-\omega_0)}{\omega_0}\right)},
\end{equation}
up to a constant phase depending on the length of the transmission line. Here
$Q$ is the resonator's quality factor, which for the circuitry in
Fig.~\ref{ResonantCircuitsFig} a) is determined by the characteristic
impedance on the transmission line $Z_0$ through $Q=\omega_0 L C^2/C_c^2
Z_0$. Since there is no dissipation in the oscillator we have
$|\Gamma(\omega)|=1$. Driving the oscillator at the bare resonance frequency
$\omega_d=\omega_0$ the phase-difference between the ground and excited state
of the qubit will be
\begin{equation}
\label{PhaseShiftEquation}
\delta\varphi_r=\varphi_r^{g}-\varphi_r^{e}=4\arctan{(Q C_Q^{\rm
max}/C)}.
\end{equation}
The phase-difference can be detected by measuring the reflected
signals in-phase and quadrature components after mixing it with
the drive.

\section{A quantum description of the quantum capacitance readout}

Above we described the quantum capacitance of the Cooper-pair box
and its use for qubit read-out in a quasiclassical manner,
treating the oscillator, transmission line, drive and detection
classically. In order to address questions about the optimal
read-out time, quantum back-action on the qubit and the quantum
efficiency of the read-out process we need a fully quantum
description of the system. In this paper we do not have the space
to go into details, which will be published
elsewhere\cite{GoranLarsChris}, but we will discuss the principles
of our model and show a fully quantum derivation of the quantum
capacitance.

The approach we chose is close to the "Quantum Network Theory"
introduced by Yurke and Denker \cite{YurkeDenker}. In
section~\ref{LagrangianSubsection} we start by writing down the
Lagrangian describing the classical dynamics of the circuit.
Through a Legendre transform we arrive at the corresponding
Hamiltonian. By stating canonical commutation relations between
our phase coordinates $\Phi_i$ and their canonical conjugate
momenta (charges) $q_i$
\begin{equation}
\left[\Phi_i,q_i\right]=i\hbar
\end{equation}
we arrive at a quantum Hamiltonian description of our circuit,
which is discussed in section~\ref{HamiltonianSubsection}. In the
relevant parameter regime we arrive at the expression for the
quantum capacitance of the Cooper-pair box. Finally in
section~\ref{HomodyneSubsection} we give an expression for the
optimal qubit read-out time using homodyne detection.

\subsection{Circuit Lagrangian}
\label{LagrangianSubsection} 
\begin{figure}[!ht]
\includegraphics[width=13cm]{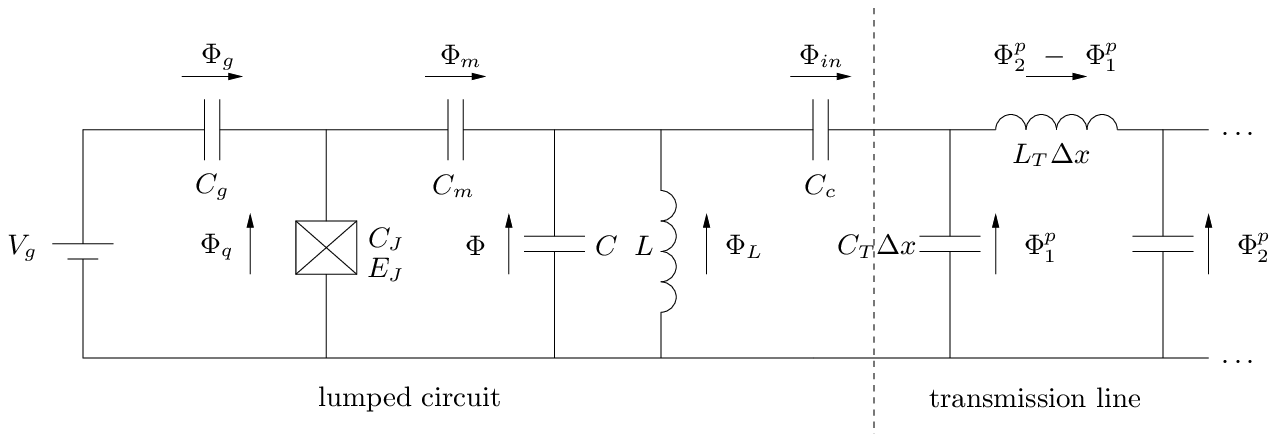}
\caption{The circuit used for measurement of the quantum
capacitance of the Cooper-pair box. It is similar to the circuit
shown in Fig.~\ref{ResonantCircuitsFig}a)+c), but the transmission
line is here modelled as a semi-infinite line of LC-circuits in
series. The phases $\Phi_i(t)=\int^t dt' V_i(t')$ across the
different circuit elements are the coordinates used in the
Lagrangian describing the dynamics of the system.}
\label{fig:circuit}
\end{figure}

The circuit for performing read-out
through the quantum capacitance is presented in figure
\ref{fig:circuit}. A Josephson charge qubit is capacitatively
coupled to a harmonic oscillator, which is coupled to a
transmission line. Through this line, all measurement on the qubit
is performed. We model the line as a semi-infinite line of
LC-circuits in series. The working point of the Josephson junction
can be chosen using the bias $V_g$. In writing down the Lagrangian
we are free to chose any quantities as our coordinates as long as
they give a full description of our circuit. Since we are treating
a system including a Josephson junction, the phases
$\Phi_i(t)=\int^t dt' V_i(t')$ across the circuit elements are
natural coordinates, as discussed by Devoret in
ref.~\onlinecite{DevoretLesHouches}. (This is in contrast to the
original work by Yurke and Denker where charges are chosen as
coordinates.)

The capacitive energy of the circuit act as the kinetic terms in
the Lagrangian\cite{WendinShumeiko2005}
\begin{equation}\label{eq:T}
T = {1\over 2}\left(C_g \dot{\Phi}^2_g +
    C_J \dot{\Phi}^2_J +
    C_m \dot{\Phi}^2_m +
    C \dot{\Phi}^2 +
    C_c \dot{\Phi}^2_{in}\right) +
    {1\over 2} \sum_{i=1}^\infty\Delta x C_T(
\dot{\Phi}^p_{i} )^2
\end{equation}
and the inductive part plays the part of potential energy
\begin{equation}\label{eq:V}
V = \frac{\Phi_L^2}{2 L} - E_J
\cos{\left(\frac{2e}{\hbar}\Phi_J\right)} + \sum_i\Delta x
\frac{(\Phi^p_{i+1} - \Phi^p_{i} )^2}{2L_T(\Delta x)^2}
\end{equation}
Applying Kirchoff's voltage law gives us the constraints
\begin{eqnarray}\label{eq:constraints}
\dot{\Phi}_g - \dot{\Phi}_J + V_g  =  0,
\Phi_J + \Phi_m - \Phi = 0,  \nonumber \\
\Phi_{in} + \Phi -  \Phi^p_1  =  0,
 \Phi - \Phi_L  =  0,
\end{eqnarray}
which gives the Lagrangian for the system
\begin{eqnarray}\label{eq:lagrangian2}
L &=& \frac{C_{qb} \dot{\Phi}^2_J}{2} + \frac{(C_{osc}+C_c)
\dot{\Phi}^2}{2} + \frac{C_c( \dot{\Phi}^p_1 )^2}{2} -
\frac{\Phi^2}{2 L} + E_J \cos{\left(\frac{2e}{\hbar}\Phi_J\right)}
- \frac{C_m}{2}\dot{\Phi}\dot{\Phi}_J -
\frac{C_c}{2}\dot{\Phi}\dot{\Phi}_1^p -
C_gV_g\dot{\Phi}_J +\nonumber\\
 &+& \sum_{i=1}^\infty \Delta x \left( \frac{C_T( \dot{\Phi}^{p}_i )^2}{2}
 -    \frac{(\Phi^{p}_{i+1} - \Phi^{p}_{i} )^2}{2L_T(\Delta x)^2}\right).
\end{eqnarray}
The capacitances in the Lagrangian are now $C_{qb} = C_J + C_g +
C_m$, and $C_{osc} = C + C_m$.

\subsection{Hamiltonian and quantum capacitance}
\label{HamiltonianSubsection}
From the Lagrangian we easily obtain
the Hamiltonian through a Legendre transform. We present the
Hamiltonian on the form
\begin{equation}
H = H_{qb} + H_{osc} + H_{TL} + H_{int},
\end{equation}
and for simplicity we assume weak coupling $C_m \ll
\{C_{osc},C_{qb}\}$ and present only the lowest order (in
 $C_m/\{C_{qb},C_{osc}\}$) terms. $H_{qb}$ contains the qubit degrees of
 freedom including the coupling of the qubit to the rest of the system
\begin{equation}\label{eq:Hq}
H_{qb} =  \frac{1}{2C_{qb}}(q_{J} + C_gV_g)^2 +
\frac{C_m}{C_{qb}C_{osc}}\left( q + q_p \right)(q_{J} + C_gV_g) -
E_J\cos{\left(\frac{2e}{\hbar}\Phi_J\right)} ,
\end{equation}
while the terms describing the oscillator, transmission line and
their interaction are
\begin{equation}
H_{osc} =  \frac{q^2}{2C_{osc}} + \frac{\Phi^2}{2L}, \ \ H_{TL} =
\frac{q_p^2}{2C_c} + \frac{1}{\Delta x} \sum_{i=1}^\infty \left(
\frac{ (q_{(i+1)}^p)^2}{2C_T} + \frac{(\Phi^p_{i+1} - \Phi^p_{i}
)^2}{2L_T}\right),  \ \ H_{int} = \frac{qq_p}{C_{osc}},
\end{equation}
where the charge operators $q, q_J, q_p$ and $q_{i}^p$ are the
conjugate momenta  to phase operators $\Phi, \Phi_{J}, \Phi^p_{1}$
and $\Phi^p_{i}$ respectively. For realizing a charge qubit the
box charging energy $E_C=e^2/2C_{qb}$ is much smaller than the
Josephson energy $E_C\gg E_J$. For relevant parameters we can then
limit the qubit charge $q_J$ to the two values $\{0,2\}$e, and we
get the usual expression for the qubit Hamiltonian in the language
of the Pauli spin matrices
\begin{equation}\label{eq:Hq3}
H_{qb} =  -\frac{E_{el}}{2}\sigma_z - \frac{E_J}{2}\sigma_x  +
2E_C\kappa\frac{q+ q_p}{e}\sigma_z + 2E_C\kappa\frac{q+
q_p}{e}(1-n_0),
\end{equation}
where we introduce the electrostatic energy-difference of the
qubit states $E_{el} = 4E_C(1-n_0)$, the dimensionless charge
$n_0=C_gV_g/e$, and the oscillator-qubit coupling coefficient
$\kappa = C_m/C_{osc} \ll 1$. The last term does not influence the
systems dynamics and may be absorbed in a small shift of $q$ and
$q_p$. Rotating the remaining two terms to the eigenbasis of the
qubit
\begin{equation}\label{eq:Hq4}
H_{qb} = \sigma_z\sqrt{16E_C^2\left(1-n_0-\kappa(q+q_p)/e\right)^2
+ E_J^2/4},
\end{equation}
we arrive at the usual charge qubit Hamiltonian with the charge
induced by the oscillator added to the induced gate charge. We now
concentrate on the case when the oscillator frequency $\omega_0 =
1/\sqrt{L C_{osc}}$ is much lower than  qubit frequency
$E_J/\hbar$. Furthermore we consider the amplitude of the
oscillators charge oscillations $q_0$ such that the induced charge
oscillations in the qubit are small $\kappa q_0/e \ll E_J/4E_C \ll
1$. In this case the qubit will follow the oscillator dynamics
adiabatically and the rates for transition between the qubit
eigenstates are negligible. Furthermore, we may Taylor expand the
qubit energy around the working point and at the charge degeneracy
$n_0=1$ we arrive at the final Hamiltonian
\begin{eqnarray}\label{eq:Hamiltonian}
H &=& -\frac{E_J}{2}\sigma_z+ \left( \frac{1}{2C_{osc}} +
\frac{4\kappa^2E_C^2}{e^2E_J}\sigma_z \right) (q+q_p)^2 +
\frac{\Phi^2}{2L}  + \frac{q_p^2}{2C_c} + \frac{1}{\Delta
x}\sum_{i = 1}^\infty \left( \frac{(q_{i+1}^p)^2}{2C_T} +
\frac{(\Phi^p_{i+1} -
  \Phi^p_i)^2}{2L_T}  \right) .
\end{eqnarray}
The qubit thus shifts the capacitative energy of the oscillator,
which in turn corresponds to adding a small extra capacitor to the
oscillator $C_\Sigma  = C_{osc} + C_Q$, where the \emph{quantum
capacitance} $C_Q$ is given by
\begin{equation}
C_Q = -\frac{2e^2C_m^2}{E_J C_{qb}^2} \sigma_z,
\label{QuantQuantCap}
\end{equation}
which is identical to the semiclassical formula in
Eq.~\ref{eq:QuantCap}. This in turn will shift the resonance
frequency of the oscillator with an amount $\delta \omega =
-\sigma_z\omega_0 C_Q/2C_{osc}$.


\subsection{Qubit read-out using homodyne detection}
\label{HomodyneSubsection}
Taking the continuum limit $\Delta x \rightarrow 0$ in
Eq.~(\ref{eq:Hamiltonian}) the solutions to the Hamiltonian for
the transmission line correspond to fields $\Phi(x\pm vt)$
propagating to the left and right with velocity $v=1/\sqrt{C_T
L_T}$. From the Hamiltonian we derive equations of motion from
which the right-propagating out-field $\Phi^{out}(t)$ and the
charge on the oscillator $q(t)$ can expressed in terms of the
left-propagating in-fields $\Phi^{in}(t)$, which is determined by
the drive. In the parameter regime relevant for qubit read-out,
where the amplitude of the qubit charge oscillations induced by
the LC-circuit is small, the qubit will follow the oscillator
adiabatically. Furthermore it is adequate to neglect third and
higher order terms in the Taylor expansion of the qubit energy in
Eq.~(\ref{eq:Hq4}). Thus we arrive at linear equations of motion
which we may solve in Fourier representation
\begin{eqnarray}
\chi(\omega) &=& \frac{i 2C_c C_\Sigma L \omega^3}
{1-(C_\Sigma+C_c)L\omega^2-i\omega C_cZ_0(1-L C_\Sigma \omega^2)}, \nonumber \\
q(\omega)+q_p(\omega) &=& \chi(\omega)  \Phi^{in}_p(\omega), \nonumber \\
\Phi^{out}(\omega) & = &
\frac{\chi(\omega)}{\chi^*(\omega)}\Phi^{in}_p(\omega)
=S(\omega)\Phi^{in}_p(\omega),
\end{eqnarray}
where $Z_0 = \sqrt{L_T/C_T}$ is the characteristic impedance of
the transmission line. Since there is no dissipation in the lumped
circuit we have $|\Phi^{out}(\omega)|=|\Phi^{in}(\omega)|$. In
this linear regime the Heisenberg equations of motion are similar
to the classical ones and for a high quality oscillator we have
$S(\omega)=\Gamma(\omega)$ as given in
Eq.~(\ref{ClassicalVoltageReflectionAmplitude}). To discuss the
quantum statistics of the qubit read-out we need a quantized
representation of the fields
\begin{eqnarray}
\label{freq_field_def_eq} \Phi^{in}(t) &=& \sqrt{\frac{\hbar
Z_0}{4\pi}} \int_{0}^\infty
\frac{d\omega}{\sqrt{\omega}}\left[a^{in}_\omega e^{-i\omega t} +
(a^{in}_\omega)^\dagger e^{i\omega t} \right], \nonumber\\
\Phi^{out}(t) &=& \sqrt{\frac{\hbar Z_0}{4\pi}} \int_{0}^\infty
\frac{d\omega}{\sqrt{\omega}}\left[S(\omega)a^{in}_\omega
e^{-i\omega t} +
S(\omega)^*(a^{in}_\omega)^\dagger e^{i\omega t} \right], \nonumber\\
q(t) &=&\sqrt{\frac{\hbar Z_0}{4\pi}}  \int_{0}^\infty
\frac{d\omega}{\sqrt{\omega}}\left[ \chi(\omega) a^{in}_\omega
e^{-i\omega t} + \chi(\omega)^*(a^{in}_\omega)^\dagger e^{i\omega
t} \right],
\end{eqnarray}
where the in-field annihilation and creation operators obey the
canonical commutation relations
\begin{equation}
[a_\omega,a^\dagger_{\omega'}] = \delta(\omega-\omega'), \ \ {\rm
and} \  [a_\omega,a_{\omega'}] = 0 .
\end{equation}

What we have achieved is a complete quantum description of the
dynamics of the oscillator and qubit in terms of the incident
field from the transmission line. We also get a full description
of the outgoing field, which is what will enter the detector. The
linearity of the equations of motion arise since we have
approximated the qubit with a state-dependent capacitance. This is
valid as long as the amplitude of the charge oscillations induced
on the qubit island are small, so that a second order Taylor
expansion of the energy in Eq.~(\ref{eq:Hq4})is enough.
Furthermore we neglect transitions between the qubit states, which
can be done for small amplitude oscillations and a low oscillator
frequency $\hbar\omega_0 \ll E_J $.

We have now the formalism needed to address questions about the
readout time which we do in the next section. We can also
calculate the backaction of the measurement process on the qubit.
This and the issue of the quantum efficiency, i. e. the relation
between the qubit dephasing rate and the measurement time will be
addressed elsewhere\cite{GoranLarsChris}.

By measuring the power of the reflected signal mixed with a local
oscillator, e.g. the drive itself, and then a local oscillator
shifted 90 degrees the in-phase and quadrature signal amplitude
can be extracted, as was done in Ref.~\onlinecite{Duty2004}. For
optimized qubit read-out it's advantageous to implement the
standard quantum optics scheme of homodyne
detection\cite{ZollerGardiner}.

To model a measurement we thus put the in-field in a coherent
Glauber state \cite{BarnettRadmore}
\begin{equation}
|\{\alpha(\omega)\}\rangle = \exp{\left(\int d\omega
[\alpha(\omega)(a^{in}_\omega)^\dag-\alpha^*(\omega)a^{in}_\omega]\right)|0\rangle},
\end{equation}
where $\alpha(\omega)$ is the Fourier-transform of our drive
signal, and $|0\rangle$ is the continuum vacuum field
$a_\omega|0\rangle=0$. We model our drive source with a narrow
($\Gamma_d \ll \omega_d$) Gaussian distribution in frequency
\begin{equation}
\alpha(\omega)=\alpha_0 \frac{\omega_d}{\Gamma_d}
\frac{e^{-(\omega-\omega_d)^2/2\Gamma_d^2}}{\sqrt{\omega}} ,
\end{equation}
where $\alpha_0$ is a dimensionless constant. For
$|t|\ll\Gamma_d^{-1}$ this gives the average electrical field
\begin{equation}
\bar{V}^{in}(x,t)=-\alpha_0 \sqrt{2\hbar Z_0}\omega_d
\sin{\left[\omega_d(t+x/v)\right]},
\end{equation}
giving on average $\Gamma^{in}_n=\alpha_0^2 \omega_d$ photons per
second sent through the transmission line by the drive.

The annihilation operator for the reflected signal has the
amplitude $a_\omega^{out}=S(\omega) a_\omega^{in} \approx
e^{-(\omega_d t-\varphi_r^{g/e})}a_\omega^{in}$ depending on the
qubit state. The signal is mixed with a strong local oscillator
with amplitude $\alpha_{LO} e^{-(\omega_d t-\varphi_{LO})}$ and
the intensity is detected. The result is then integrated for a
time $T$. The intensity is given by the number of photons incident
on the detector
\begin{equation}\label{eq:intensity}
N(\omega,T) = \int_0^T dt \; b^\dagger(t)b(t)
\end{equation}
where the field at the detector is
\begin{equation}
b(t) = r\alpha_{LO}(t) + rv(t) + t a^{out}(t),
\end{equation}
here $r \ll 1$  denotes the small reflection coefficient of the
mixer and $t$ is the corresponding transmission coefficient, and
$v(t)$ is the vaccuum-field of the idle mixer port. (The mixer is
a beam-splitter in the quantum optics case.) We now assume that
$|r \alpha_{LO}(t)| \gg 1$ is large and neglect second order
contributions in $v(t)$ and $a(t)$ to equation
(\ref{eq:intensity}), arriving at the average photon number at the
detector
\begin{equation}
\langle N^{g/e} \rangle  = T\left(r^2 \alpha_{LO}^2 + 2tr
\alpha_{LO} \alpha_0^{in}
\cos{(\varphi_r^{g/e}-\varphi_{LO})}\right),
\end{equation}
where $\alpha_0^{in}$ is amplitude of the in-field. The first term
is a pure local oscillator term and contains no information, while
the second term is maximized choosing
$\varphi_{LO}=\frac{\varphi_r^g+\varphi_r^e}{2}-\frac{\pi}{2}$.
The two probability distributions of the number of detected
photons will be separated by two times the variance after the
measurement time
\begin{equation}
T_{ms}=\frac{1}{\Gamma_n^{in}}\frac{1}{4
\sin^2{\left[\frac{\varphi_r^g-\varphi_r^e}{2}\right]}} ,
\end{equation}
where $\Gamma_n^{in}$ is rate of photons sent into the
transmission line by the drive. For not too low quality factor of
the oscillator we can use the Breit-Wigner approximation for
$\chi(\omega)$ leading to the expression for the phase-shift in
Eq.~(\ref{PhaseShiftEquation}). Thus we find for the measurement
time
\begin{equation}
T_{ms}=\frac{1}{\Gamma_n^{in}}\frac{1}{4
\sin^2{\left[2\arctan{x}\right]}}=\frac{(x^{-1}+x)^2}{16\Gamma_n^{in}},
\ \ x=\frac{Q C_Q^{\rm{max}}}{C} .
\end{equation}
From the condition that the drive should induce only small
oscillations of the qubit charge we arrive at the following bound
for the drive strength
\begin{equation}
\Gamma_n^{in} < \frac{E_J}{16\hbar}(x+x^{-1})
\end{equation}
giving a lower bound on measurement time
\begin{equation}
T_{ms} > \frac{\hbar}{E_J} (x+x^{-1}) ,
\end{equation}
indicating that the measurement time must be larger than
$\hbar/E_J$, which is not very restrictive. Due to the oscillator
"ring-up" time the measurement time is further limited by $T_{ms}
> Q/\omega_0 = x C/C_Q\omega_0$. For a fixed $\omega_0$ this
indicates that the regime $x \ll 1$ is advantageous. Comparing
these two inequalities we find the shortest measurement time for
\begin{equation}
x_{opt}^2=\frac{C_Q}{C}\frac{\hbar\omega_0}{E_J}, \ \
Q_{opt}=\sqrt{\frac{C}{C_Q}\frac{\hbar\omega_0}{E_J}}, \ {\rm and}
\ T_{ms} > \frac{Q_{opt}}{\omega_0} ,
\end{equation}
implying that a low $Q$ is clearly an interesting regime. For low
$Q$ the Breit-Wigner approximation of $\chi(\omega)$ breaks down,
and so does the simple estimates of the measurement, but the
formalism developed here is still applicable using the full
expressions. An optimization including the measurement induced
back-action on the qubit will be discussed in
Ref.~\onlinecite{GoranLarsChris}.


\subsection{A comparison with dispersive readout using a non-adiabatic oscillator}
In an experiment at Yale university\cite{WallraffReadOutPRL2005} a
charge qubit coupled capacitatively to a microstrip cavity was
read out by sending microwaves through the cavity. The qubit state
influences the resonance frequency and thus the phase-shift of the
transmitted signal. This phase difference was then detected in a
similar fashion as described above.

The main difference compared to what was discussed above
is that the cavity resonance frequency (5.4 GHz) was of
the same order of magnitude as the qubit frequency (4.3 GHz). In
this regime it is appropriate to use the rotating wave
approximation and the system dynamics is described by the
Jaynes-Cummings hamiltonian\cite{Blais2004}.

For a comparison we start with the Hamiltonian of a qubit coupled
transversely to a harmonic oscillator
\begin{equation}
H=-\frac{\hbar\omega_{qb}}{2} \sigma_z + i g \sigma_x (a^\dag-a) +
\hbar\omega_{osc}\left(a^\dag a+\frac{1}{2}\right),
\end{equation}
applicable for a charge qubit at charge degeneracy, coupled
capacitatively to the harmonic oscillator. Performing a
straightforward second order perturbation expansion in the
coupling term we find the renormalized spectrum
\begin{eqnarray}
E_{n\uparrow} & = & \hbar\omega_{osc}\left(n+\frac{1}{2}\right) -
\frac{\hbar\omega_{qb}}{2} - g^2 \left[
\frac{n}{\hbar\omega_{qb}-\hbar\omega_{osc}} +
\frac{n+1}{\hbar\omega_{qb}+\hbar\omega_{osc}} \right] \ {\rm and}
\nonumber
\\
E_{n\downarrow} & = & \hbar\omega_{osc}\left(n+\frac{1}{2}\right)
+ \frac{\hbar\omega_{qb}}{2} + g^2 \left[
\frac{n}{\hbar\omega_{qb}+\hbar\omega_{osc}} +
\frac{n+1}{\hbar\omega_{qb}-\hbar\omega_{osc}} \right] ,
\end{eqnarray}
where $n$ indicates the number of photons in the oscillator and
$\uparrow/\downarrow$ the qubit in the ground/excited state for
the unperturbed state $(g \rightarrow 0)$. The spectrum is formed
by two equidistant sets of energy levels, where the effective
qubit-dependent frequency shift of the oscillator amounts to
\begin{eqnarray}
\delta\omega = - \sigma_z g^2 \left[
\frac{1}{\hbar\omega_{qb}-\hbar\omega_{osc}} +
\frac{1}{\hbar\omega_{qb}+\hbar\omega_{osc}} \right],
\end{eqnarray}
being negative for the qubit in the ground state. In the regime
$|\omega_{qb} - \omega_{osc}| \ll \omega_{qb}$ only the first term
contributes and we arrive at the Jaynes-Cummings result
$\delta\omega=-\sigma_z g^2/(\hbar\omega_{qb}-\hbar\omega_{osc})$.
In the adiabatic regime $\omega_{osc} \ll \omega_{qb}$ we can
neglect the terms $\pm\hbar\omega_{osc}$ in the denominators
giving the result below Eq.~(\ref{QuantQuantCap}),
$\delta\omega=-\sigma_z 2 g^2/\hbar\omega_{qb}$. Thus we see that
the frequency shift given by the Jaynes-Cummings hamiltonian can
be described on the same footing as the one given by the quantum
capacitance.

\section{Flux measurement}

The 2-level quantum states of the persistent-current flux qubit
(Fig. 1) are characterized by different directions of the
persistent currents circulating in the qubit loop, hence different
directions of the induced magnetic flux. The flux qubit readout is
based on the detection of the induced flux or direct measurement
of the persistent currents. The latter method is also relevant for
charge qubits with loop-shape electrodes (e.g.
quantronium\cite{Vion2002}) where the intensity of the induced
flux is too small to be detectable while detection of the
persistent current is possible. In flux qubits with larger
Josephson junctions, persistent currents are large, and the
measurement of flux is not that difficult
\cite{Chiorescu2003,NTT2003,Ilichev2003,Lupascu2005}. The
experimental measurement
setup is sketched in Fig. 1, left picture: the qubit loop is
inductively coupled to a dc-SQUID connected to a current source.
Direction of the persistent current in the qubit loop affects the
magnetic flux threading the SQUID and thus affects the SQUID
critical current as well as its plasma frequency. This allows one
to make the two types of measurements, by probing the dc and the
ac properties of the measurement SQUID. In the first case, the
critical current is measured by applying a dc current slowly
increasing with time, and detecting the value of the critical
current when the SQUID switches to the resistive branch (threshold
detection), repeating the measurement to create a histogram of the
events. In the second case, an ac current is applied and the phase
shift of the reflected signal is measured. The latter method is
also possible to realize using a linear LC-oscillator instead of
the dc SQUID\cite{Ilichev2003}.

\begin{figure}[!ht]
\label{squi}
\includegraphics[width=8cm]{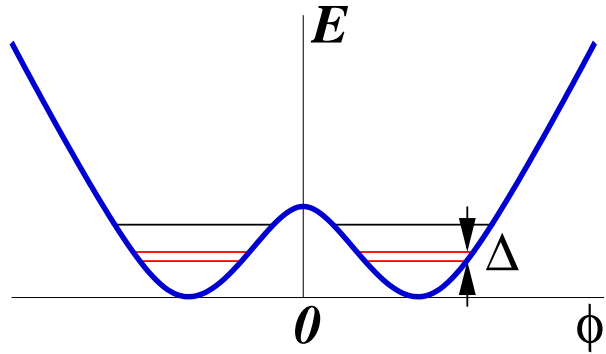}
\caption{Double-well potential and energy levels of the flux qubit
($f_q=\pi$). }
\end{figure}

To quantitatively analyze the circuit (see e.g. Refs.
\onlinecite{Burkard2004,WendinShumeiko2005,Bertet2005}), let us for
simplicity consider the single junction flux qubit, the analysis also applies
to the experimental three junction qubits. The circuit Hamiltonian consists
of the Hamiltonian $H_q$ of the qubit loop, the Hamiltonian of the SQUID,
$H_S$, and the Hamiltonian of the inductive coupling, $H_{int}$. The qubit
Hamiltonian has the form given in Eq. (\ref{HQsquid}), here we modify the
notations,
\begin{equation}\label{Hq}
H_q = E_{Cq} n^2_q - E_{Jq}\,\cos\phi_q + E_{Lq}\tilde\phi_q^2,
\end{equation}
where $E_{Cq}= (2e)^2/2C_q$ is the charging energy of the qubit junction,
$E_{Jq}$ is the Josephson energy, $E_{Lq}= (\Phi_0/2\pi)^2(1/2L_q)$ is the
inductive energy of the qubit loop, and $\tilde\phi_q = \phi_q - f_q$ is the
induced flux, $f_q$ is the biasing flux. The SQUID Hamiltonian is given by
equation,
\begin{equation} \label{Hsquid}
H_s = \sum_i^2 \,(E_{Cs} n^2_{i} - E_{Js}\,\cos\phi_i) \, +\, E_{Ls}\phi_s^2
+ {\hbar\over 4e}I(t)(\phi_1-\phi_2),
\end{equation}
where the induced flux is $\phi_s = \phi_1 + \phi_2 - f_s$, and
$I(t)$ is a (non)stationary current bias. The interaction term has
the form
\begin{equation}\label{Hint}
H_{int} = E_{M}\tilde\phi_q \phi_s,
\end{equation}
where the interaction energy is determined by the mutual
inductance $M$, $E_M = (\Phi_0/2\pi)^2 (M/L_qL_s)$.

Now we truncate the Hilbert space of the circuit Hamiltonian to
the lowest energy states, which include the two almost degenerate
(for $E_{Cq} \ll E_{Jq}$ and at $f_q\approx \pi$) lowest energy
states in the potential wells of the qubit potential energy,
Fig. 6, and the ground state of the SQUID. Then the qubit
Hamiltonian takes the form, in the eigenbasis of the non-coupled
wells
\begin{equation}\label{Hq2}
H_q = -{1\over 2}(\epsilon\sigma_z + \Delta\sigma_x).
\end{equation}
Here $\epsilon(f_q)$ is the energy level difference in the wells proportional
to $f_q-\pi$ , and $\Delta$ is the energy level splitting due to the
macroscopic tunneling between the wells. The truncated interaction term takes
the form,
\begin{equation}\label{Hint2}
H_{int} = E_{M}\phi_0 \phi_s\sigma_z,
\end{equation}
where $\phi_0$ is the half distance between the minima of the
potential energy. The off-diagonal term in the interaction is
neglected since it is exponentially small.

To truncate the SQUID Hamiltonian, we introduce new variables, $\phi_\pm =
(1/2)(\phi_1 \pm \phi_2)$, and  $\phi_s = 2\phi_+ - f_s$,
\begin{equation}\label{Hsquid2}
H_s = {1\over 2}E_{Cs} n^2_{-} + 2E_{Cs} n^2_{s} - 2E_{Js}\,\cos\phi_- \cos
\left({f_s + \phi_s\over 2}\right) \, +\, E_{Ls}\phi_s^2 + {\hbar\over
2e}I(t)\phi_-.
\end{equation}
Now we assume that the inductive energy is sufficiently large,
$E_{Ls} \gg E_{Js}$, to provide small fluctuation of the induced
flux, $\phi_s \ll 1$. This allows us to expand the cosine term;
then keeping the first order term with respect to $\phi_s$
(non-vanishing for $f_s \neq 0$), and taking into account the
interaction term \ref{Hint2}, we write the $\phi_s$-dependent part
of the total Hamiltonian on the form,
\begin{equation}\label{Hphis}
H(\phi_s) = 2E_{Cs} n^2_{s} + \left( E_{Js}\,\cos\phi_- \sin
\left({f_s\over 2}\right) + E_{M}\phi_0(f_q)\sigma_z \right)
\phi_s \, +\, E_{Ls}\phi_s^2.
\end{equation}
This is the linear oscillator shifted from the origin, the shift being
proportional to the induced flux in the qubit loop. Making projection on the
ground state of this Hamiltonian, we arrive at the non-trivial,
qubit-dependent part having the form,
\begin{equation}\label{lambda}
H(\phi_s) \rightarrow  - 2E_{Js}\lambda
\sin\left({f_s\over2}\right) \, \sigma_z \cos\phi_-\, ,\;\;\;
\lambda = {E_M\phi_0\over 8 E_{Ls}}.
\end{equation}
Combining this with the rest of the SQUID Hamiltonian and the
truncated qubit Hamiltonian, we finally get:
\begin{equation}\label{Hfinal}
H = -{1\over 2}(\epsilon\sigma_z + \Delta\sigma_x) + {1\over 2}E_{Cs} n^2_{-}
- 2E_{Js} \left[
 \cos\left({f_s\over 2}\right) + \lambda\sin \left({f_s\over
2}\right) \sigma_z \right]\,\cos\phi_- + {\hbar\over 2e}I(t)\phi_-.
\end{equation}
For the 3-junction qubit\cite{Mooij1999} the coupling constant $\lambda$ in
Eq. (\ref{lambda}) acquires an additional factor\cite{Mattias}
$E_{Jq}/E_{Lq}$, which results from tracing out the plasma mode in the qubit
loop. This mode does not form the qubit in the 3-junction circuit (in
contrast to the single-junction qubit), but this mode is an auxiliary one
connecting the qubit to the outside world, and it is eliminated similar to
the SQUID variable $\phi_+$ in Eqs. (\ref{Hphis}) and (\ref{lambda}).

The Hamiltonian (\ref{Hfinal}) describes a flux qubit directly
coupled via an effective coupling constant $\lambda$ to a
non-linear Josephson oscillator. The coupling affects the
Josephson energy of the oscillator;  hence the critical bias
current, i.e. the the magnitude of the bias dc current at which
the oscillator switches to the dissipative regime. Quantitatively,
these critical current values for the 3-junction qubit are,
\begin{equation}\label{}
{\hbar\over 2e}I_c = 2E_{Js} \left(
 \cos\left(f_s/ 2\right) \pm \lambda{ E_{Jq}\over E_{Lq}}
 \sin \left(f_s/2\right)\right).
\end{equation}

The advantage of this method is that the measurement circuit can be
disconnected during the time period between the measurements by
switching off the flux through the SQUID, $f_s=0$, thus enhancing
the decoherence time of the qubit.
The disadvantage of the method is
that for slow readout (low Josephson plasma frequency of the
SQUID compared to the qubit frequency) the switching current
depends on the average value of the induced flux, $\langle
\sigma_z \rangle$, which equals zero at the degeneracy point,
$\epsilon=0$. Thus measurement can only be performed by departing
from the degeneracy point, which is undesirable due to enhanced
decoherence. This difficulty can be solved by probing the qubit
quantum inductance, which is analogous to the quantum capacitance
measurement for charge qubits.

\section{Inductance measurement}
\begin{figure}[!ht]
\label{LC}
\includegraphics[width=10cm]{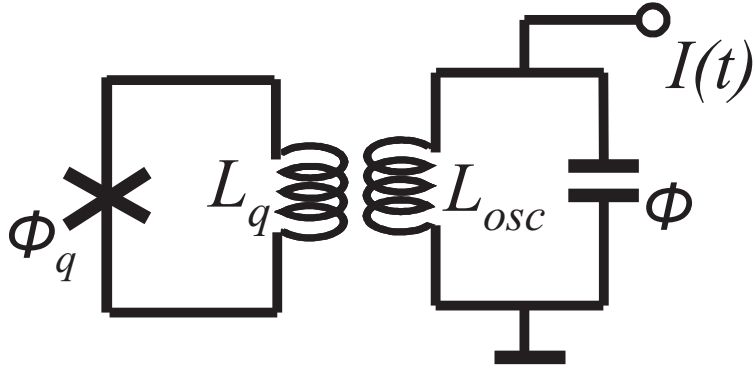}
\caption{Single-contact flux qubit inductively coupled to a linear
oscillator. }
\end{figure}
Consider first a simpler circuit with a linear LC-oscillator
replacing the dc SQUID (Fig. 7). Such a device, rf-SQUID
inductively coupled to a linear oscillator is a classical device
employed for many years for precise measurement of magnetic
field\cite{Barone}. The principle of operation is based on the
magnetic field dependence of the Josephson inductance, which
affects the resonance frequency of the oscillator probed by an
external rf signal. The classical Hamiltonian for this circuit, again
assuming for simplicity a single Josephson junction in the qubit
loop, has the form,
\begin{equation}\label{Hclassic}
H = E_{Cq} n^2_q - E_{Jq}\,\cos\phi_q + E_{Lq}\tilde\phi_q^2 + E_M
\tilde\phi_q\phi \,+ \,E_{Cosc} n^2 \, +\, E_{Losc}\phi^2 +
{\hbar\over 2e}I(t)\phi.
\end{equation}
Neglecting the junction capacitance energy, and expanding the
qubit potential energy near the minimum point
$\tilde\phi_{q0}(f_q)$,
\begin{equation}\label{}
- E_{Jq}\,\cos\phi_q + E_{Lq}\tilde\phi_q^2 \equiv U(\tilde\phi_q) =
U(\tilde\phi_{q0})+ {1\over 2}U^{\prime\prime}(\tilde\phi_{q0}) (\tilde\phi_q
- \tilde\phi_{q0})^2,
\end{equation}
we define an effective inductance of the qubit circuit $L_q^{eff}$
via the relation
\begin{equation}\label{}
{1\over 2}U^{\prime\prime}(\tilde\phi_{q0})=
(\Phi_0/2\pi)^2/2L_q^{eff}(f_q)= E_{Lq}^{eff}(f_q) .
\end{equation}
After having diagonalized the total potential energy, we obtain
the shift of the oscillator inductive energy due to coupling to
the qubit,
\begin{equation}\label{shiftclassic}
\tilde E_{Losc} = E_{Losc} - {E_{M}^2\over 4E_{Lq}^{eff}(f_q)}.
\end{equation}
This gives rise to a shift of the oscillator resonance frequency
depending on the magnetic flux through the qubit loop, which is
probed with an external rf signal $I(t)$.

A similar measurement procedure also applies to the quantum regime.
The quantum Hamiltonian for the same circuit has the form, taking
into account Eqs. (\ref{Hq2}) and (\ref{Hint2}),
\begin{equation}\label{Hqosc}
H= -{1\over 2}(\epsilon\sigma_z + \Delta\sigma_x)+
E_{M}\phi_0\phi\sigma_z + E_{Cosc} n^2 \, +\, E_{Losc}\phi^2 +
{\hbar\over 2e}I(t)\phi.
\end{equation}
For a slow oscillator and weak coupling, the Hamiltonian can be
rotated to a qubit eigenbasis, and expanded with respect to the
coupling term,
\begin{equation}\label{Hqosc2}
H= -{1\over 2}\varepsilon\sigma_z - {(E_{M}\phi_0)^2\over
\varepsilon}\,\phi^2\sigma_z + E_{Cosc} n^2 \, +\, E_{Losc}\phi^2
+ {\hbar\over 2e}I(t)\phi
\end{equation}
(here the term linear in $\phi$ is omitted since it only produces
a non-essential small shift of the oscillator coordinate;
$\varepsilon = \sqrt{\epsilon^2 + \Delta^2}$). The second term in
Eq. (\ref{Hqosc2}) provides the shift of the oscillator inductive
energy depending on the qubit state,
\begin{equation}\label{}
\tilde E_{Losc} = E_{Losc} - {E_{M}^2\phi_0^2\over
\varepsilon}\,\sigma_z.
\end{equation}
Comparing this quantum result with the classical equation
(\ref{shiftclassic}), we are able to identify the quantum
inductance of the qubit,
\begin{equation}\label{}
L_Q =  \left({2\pi\over \Phi_0}\right)^2{2\phi_0^2\over
\varepsilon}\,\sigma_z.
\end{equation}
The quantum inductance is inversely proportional to the qubit
level splitting, similar to the quantum capacitance of charge
qubits, and it approaches its maximum value at the degeneracy point.

This conclusion also applies to the 3-junction flux
qubit\cite{Mooij1999}; the only difference is due to the
suppressed coupling of the qubit to the outside world discussed in
the previous section below Eq. (\ref{Hfinal}). In this case, an
additional small factor, $E_{Jq}/E_{Lq}$, appears in the the
coupling term in Eq. (\ref{Hqosc}), giving rise to the following
equation for the quantum inductance of the 3-junction qubit,
\begin{equation}\label{}
L_Q =  \left({2\pi\over \Phi_0}\right)^2\left({E_{Jq}\over
E_{Lq}}\right)^2{2\phi_0^2\over \varepsilon}\,\sigma_z.
\end{equation}

Proceeding with the case of the readout dc SQUID, Eq.
(\ref{Hfinal}), we find that this case seems to be qualitatively
different from the LC-oscillator readout: the qubit-meter coupling
is non-linear. Obviously, this results from the fact that the
qubit is not directly coupled to the readout $\phi_-$-oscillator,
but rather via an intermediate $\phi_s$-oscillator, the elimination of
which results in the non-linear coupling.

One way to solve the problem is to displace the
$\phi_-$-oscillator by applying a constant current bias, $I(t) =
I_0 +I_1(t)$. Then expanding the potential energy around the
minimum point, $\phi_- = \bar\phi_- + \theta$, where $\bar\phi_-$
satisfies the equation, $2E_{Js}\cos(f_s/2)\sin\bar\phi_- = -
(\hbar/2e)I_0$, we arrive, in the linear approximation, at a
Hamiltonian similar to Eq. (\ref{Hqosc}),
\begin{equation}\label{Hfinal2}
H = -{1\over 2}(\epsilon\sigma_z + \Delta\sigma_x) + {1\over 2}E_{Cs} n^2_{-}
+ E_{Js}\cos\bar\phi_-\cos(f_s/2)\theta^2 + \tilde\lambda
E_{Js}\theta\sigma_z  + {\hbar\over 2e}I_1(t)\theta,
\end{equation}
\begin{equation}\label{}
\tilde\lambda = \lambda \,{\hbar I_0 E_{Jq}\over 2eE_{Lq}}
 \tan \left(f_s/2\right).
\end{equation}
Another solution would be to access directly the $\phi_s$-oscillator linearly
coupled to the qubit. This can be done, for example, by driving a bias flux
through the SQUID, $f_s(t)$. For small variation of phases in
Eq.(\ref{Hsquid2}), the oscillators decouple, and the relevant part of the
Hamiltonian, taking into account Eqs. (\ref{Hq2}) and (\ref{Hint2}),
approaches a form similar to  Eq. (\ref{Hqosc}),
\begin{equation}\label{}
 H = -{1\over 2}(\epsilon\sigma_z + \Delta\sigma_x)+
 + 2E_{Cs} n^2_{s}  + E_{Js}\phi_s^2 + {E_ME_{Jq}\phi_0
 \over E_{Lq}}\,\phi_s\sigma_z + 2E_{Js}f_s(t)\phi_s.
\end{equation}
Finally, the $\phi_s$-oscillator  may be accessed by using an
asymmetric SQUID, with different inductances of the left and right
legs (Fig. 1), $L_1\neq L_2$, $L_1+L_2 = L_s$. In this case,
the coupling of the SQUID to an external current source in Eq.
(\ref{Hsquid}) becomes modified to
\begin{equation}\label{}
{\hbar\over 4e}I(t)\left({L_2\over L_s}\,\phi_1 - {L_1\over
L_s}\,\phi_2\right),
\end{equation}
which results in direct coupling of the probing current to the
$\phi_s$-oscillator. At zero flux bias, $f_s=0$, the two
$\phi_\pm$-oscillators do not interact, and the relevant part of
the Hamiltonian reads,
\begin{equation}\label{}
H= -{1\over 2}(\epsilon\sigma_z + \Delta\sigma_x)+
 2E_{Cs} n^2_s \,
+\, (E_{Ls}+E_{Js})\phi^2_s \,+\, {E_{M}E_{Jq}\phi_0\over
E_{Lq}}\,\phi_s\sigma_z \,+\, {\hbar\over 2e}I(t)\,{L_2-L_1\over
L_s}\,\phi_s.
\end{equation}
Thus there are several ways to employ the dc SQUID for dispersive
measurement of the qubit quantum inductance. Note, however, that
in the latter case the inductive energy of the SQUID plays a role, and
for  small SQUID inductance ($E_{Ls}\gg E_{Js}$) the oscillator
frequency may become large, violating the adiabatic regime assumed in
the derivation.

\section{Concluding remarks}

In this paper we have outlined some practical schemes for
capacitive and inductive readout, detecting the state of a qubit
by reflecting microwaves from an oscillator circuit, the phase
shift measuring the changes in charge or magnetic flux induced by
a qubit, allowing to distinguish between the different states
$\ket{0}$ and $\ket{1}$ of the qubit. In particular we focussed
attention on the single-Cooper-pair box (SCB), the effective
capacitance of which can be defined as the derivative of the
induced charge with respect to gate voltage. In addition to the
geometric capacitance, there is the quantum capacitance due to the
level dispersion at the anti-crossing caused by the Josephson
coupling. We described the process of reflection of quantized
radiation and derived expressions for the shortest measurement
time needed to resolve the qubit states, suggesting that a low Q
should be advantageous for weak back action fast readout.

\section{Acknowledgment}
This work was supported in part by the European Commission through projects
FP5-39083-SQUBIT-2 and FP6-015708 EuroSQIP of the IST Priority
(Disclaimer: www.eurosqip.org), and by the Swedish Research Council.

\end{document}